# Temporal behavior of microwave sheath-voltage combination plasma

Satyananda Kar[1]*, Hiroyuki Kousaka[1] and Laxminarayan. L. Raja[2]

[1]*Department of Mechanical Science and Engineering, Nagoya University, Furo-cho, Chikusa-ku, Nagoya - 464 8603, Japan*

[2]*Department of Aerospace Engineering and Engineering Mechanics, The University of Texas at Austin, Austin, Texas - 78712 1221, USA*

*babuliphy@gmail.com

Microwave sheath-Voltage combination Plasma (MVP) is a high density plasma source and can be used as a suitable plasma processing device (e.g., ionized physical vapor deposition). In the present report, the temporal behavior of an argon MVP sustained along a direct-current biased Ti rod is investigated. Two plasma modes are observed, one is an "oxidized state" (OS) at the early time of the microwave plasma and the other is "ionized sputter state" (ISS) at the later times. Transition of the plasma from OS to ISS, results a prominent change in the visible color of the plasma, resulting from a significant increase in the plasma density, as measured by a Langmuir probe. In the OS, plasma is dominated by Ar ions and the density is order $10^{11}$ cm$^{-3}$. In the ISS, metal ions from the Ti rod contribute significantly to the ion composition and higher density plasma ($10^{12}$ cm$^{-3}$) is produced. Nearly uniform high density plasma along the length of the Ti rod is produced at very low input microwave powers (around 30 W). Optical emission spectroscopy measurements confirm the presence of sputtered Ti ions and Ti neutrals in the ISS.

## I. INTRODUCTION

Various high density plasma sources have been developed for plasma enhanced surface treatments such as surface cleaning, surface activation, surface coating [1] and fabrication of ultra-large-scale integrated (ULSI) circuits [2-4]. In these fields, development of high density plasma sources has been motivated mainly by requirements of high processing speed. Surface wave excited plasma (SWP) is one such high-density plasma source which, generates over-dense (greater than the cutoff plasma density, i.e., $7.6\times10^{10}$ cm$^{-3}$ for 2.45 GHz microwave) plasmas in large diameter geometries. SWP sources can be divided [5] into dielectric-bounded type (surface waves propagate along the dielectric and plasma interfaces) and metal-bounded type, employing microwave sheath-voltage combination plasma (MVP), where surface waves propagate along the plasma and sheath interfaces.

In general, dielectric-bounded SWP sources employ dielectrics such as a quartz plate [3-4, 6] or a quartz tube [7-10], along which high density plasmas are sustained by electromagnetic waves travelling along plasma-dielectric surface boundaries [10-21]. Dielectric-bounded SWP sources have never been used for sputtering, because they are not generated near and along a metal target as in conventional magnetron sputtering sources [1].

In 2005, Kousaka et. al. [22] at Nagoya University have reported a new SWP source which can generate over-dense plasma along the surface of a negatively biased metal rod, called metal-antenna surface wave excited plasma (MASWP). The source was subsequently, termed as MVP [23]. In MVP sources, microwave plasma is produced along a dielectric-free metal antenna and the surface waves propagate along the plasma-sheath interface [22-27]. The MVP source is expected to generate uniformly high-density plasma along three dimensional surfaces of a metal object [26]. In our previous work [5], a better design of MVP source was reported. In that new design, the antenna was extended from vacuum chamber to waveguide and over-dense plasma ($10^{11}$ $cm^{-3}$) was produced by very low input microwave powers (around 30 W). Here, the temporal behavior of a MVP source and generation of high-density plasma for low input microwave powers are reported.

## II. EXPERIMENTAL SET-UP

The schematic diagram of the experimental setup is shown in Fig. 1. The experiment was performed in a grounded cylindrical chamber of stainless steel (SUS304, JIS) with an inner diameter of 146 mm and a length of 500 mm. The schematic shows 5 components: (1) a vacuum chamber (anode/ground), (2) a coaxial electromagnetic wave-guide (WX-39D) for the microwave power feed-in, (3) a central conductor or antenna (titanium rod, 18 mm in diameter and 250 mm in length inside the vacuum chamber) which is set to a negative direct-current (DC) bias (cathode), (4) a dielectric window (quartz with dielectric constant $\epsilon_r \approx 3.78$ and thickness of 10 mm) which seals the vacuum chamber to the coaxial wave-guide and couples the electromagnetic wave, and (5) a Langmuir probe (cylindrical probe of 8 mm length and 0.2 mm diameter) for measurement of plasma parameters. The chamber was evacuated by a rotary pump and the base pressure was about 0.5 Pa. The working gas was argon at a pressure of 50 Pa.

The important part in our experimental system is the antenna. The antenna was long enough, i.e., 250 mm length inside the vacuum chamber and was extended from vacuum chamber to the waveguide. The old design [25-26] was used a special quartz window interrupting the antenna (detrimental to the microwave power coupling) and needed supplementary DC connection feed through (2 port design). This new design feeds the DC and the microwave power from one port only. The waveguide part of the antenna was cooled by water during the experiments.

First, the cathode/antenna was set to a negative voltage (-300 V), so a low density plasma can form with an ion sheath layer along the antenna. Then the microwave was fed into the coaxial wave-guide from the right side of the system. Now the input microwave (TM mode) power could propagate easily along the negative biased antenna and surface wave excited plasma generated along the antenna inside the chamber. The input microwave power was 30 W to 50 W. The surface wave excited plasma columns were photographed by a camera through a side viewport. A rf compensated Langmuir probe (SmartProbe, Scientific Systems) was inserted in the axial direction (opposite direction of antenna and the gap between the Langmuir probe and antenna was 26 mm), equipped with an automatic linear drive to measure the axial variation of plasma density.

The signals of optical emission spectra from the plasma were transferred by optical fiber coupled to the spectrometer (HR4000CG-UV-NIR, focal length of 101 mm) equipped with a 300 lines/nm diffraction grating (type HC-1 Landis ). The optical resolution is 0.75 nm FWHM. Light was collected and transferred to the entrance slit (5 μm) by a multiplex fiber bundle. The spectrometer was scanned in the wavelength range 200-1100 nm. Then a 8μm×200μm pixel size (3648 pixel number) CCD (Toshiba TCD1304AP linear CCD array) collected the dispersed signal from grating. The CCD signals were sent to the computer using the SpectraSuit software for data acquisition. Plasus SpecLine software was used for peak marking and spectral analysis.

## III. RESULTS AND DISCUSSIONS

Figure 2 shows the surface wave excited plasma columns along the antenna at various times from the start of the microwave plasma generation (7, 11 and 30 minutes). It is clearly seen that the length of plasma column increases with increase in time and also color changes. Here the gas pressure was 50 Pa, the applied DC bias was -300 V and the input microwave power ($P_{in}$) was 50 W. The discharge current through the DC biased antenna was approximately same up to 10 minutes (Fig. 2(a)). After 10 minutes, plasma color starts changing and the discharge current increases (Fig. 2(b)) and after around 20 minutes, the discharge current saturates (Fig. 2(c)) as well as plasma column. Figure 2(d) shows the close-up view of the plasma at 30 minutes. Figure 3 shows the time variation of the discharge current at DC biased antenna for various input microwave powers.

From Fig. 2, plasma color change after the initial transient that lasts 10 to 15 minutes is evidence that a significant change in the composition of the plasma occurs. We hypothesize that the color/composition change results from a high rate of sputtering of the Ti antenna. Normally the sputter yield for Ti in an argon plasma for ion energy in the range 0 to 300 eV is 0 to 0.307 (threshold energy of Ti is 25.4 eV) [28]. The sputter yield of compounds (with an oxidized surface) is generally lower than the sputter yield of metals [29]. We can therefore define two states. At the early time of the MVP (Fig. 2(a)), the antenna is covered by a thin oxidized layer that has a low sputter yield and consequently the plasma is negligibly contaminated by antenna material. We term this the "oxidized state" (OS). At later times (Fig. 2(b) & 2(c)), the oxide layer is eventually removed by the slow oxide sputtering process to expose the underlying pristine metal surface. The metal itself has a high sputter yield causing a large flux of metal atoms to enter the plasma and thereby changing the composition significantly. We term this the "ionized sputter state" (ISS).

Figure 4 shows the axial distributions of ion density for OS and ISS. These measurements are taken from the left side of the system (300 mm is the antenna zero position corresponding to the microwave inlet). For OS conditions (7 minutes), the density is around $1\times10^{11}$ $cm^{-3}$. For ISS conditions (15, 20 and 30 minutes), an over-dense plasma ($\geq 1\times10^{12}$ $cm^{-3}$) is generated, which is much greater than the cutoff plasma density (i.e., $7.6\times10^{10}$ $cm^{-3}$ for 2.45 GHz microwave). Clearly, under ISS conditions the ion density distributions are more uniform in length and higher than OS.

Figure 5 shows the axial ion density and electron temperature distributions for various microwave input power $P_{in}$ for ISS conditions (after the discharge current saturates). Here the discharge currents through the antenna were varied in the span from 0.53 A to 0.77 A for input microwave powers from 30 W to 50 W. In Fig. 4(a) the length of plasma column and ion density increases with increase in $P_{in}$. Figure 4(b) shows the electron temperature distribution for various $P_{in}$. For the lowest power of 30 W the peak temperature is about 2 eV and it increases to about 4 eV for the high power conditions.

Optical emission spectroscopy was used to provide insights into the composition of the plasma for the OS and ISS conditions. Figure 6 shows a survey spectrum of the plasma over a wavelength range from 350-1000 nm. The top panel (Fig. 6(a)) corresponds to the OS condition and the bottom panel (Fig. 6(b)) corresponds to the ISS conditions for the same case. The applied DC bias to the antenna (-300 V), the gas pressure (50 Pa) and $P_{in}$ (30 W) were fixed. The axial position was 120 mm away from the antenna zero position. For OS conditions (Fig. 6(a)), the spectrum is dominated by Ar atoms (ArI) and Ar ions (ArII) in the wavelength range of 675-1000 nm and few oxygen (OI and OII) lines also in the same wavelength range. There is no indication of Ti lines for the OS condition. For ISS conditions (Fig. 6(b)), the spectrum shows the emergence of additional lines at lower wavelengths. These lines correspond to ArI, ArII, and OII. In addition lines corresponding to Ti species: Ti atom (TiI), singled ionized Ti (TiII), and doubly ionized Ti (TiIII) appear at 430.75, 466.35, 467.5, 468.7, 609.2, 613.2 and 660.56 nm.

These results from emission spectra clearly buttress our hypothesis that the OS is a pre-sputter state where negligible amounts of material from the metal antenna is introduced into the plasma and therefore the plasma is a nearly pure argon plasma. The oxidized metal layer acts as a protective layer that prevents rapid sputtering of the antenna. Once the oxide layer is removed (by sputtering, albeit at a very slow rate) the metal is exposed which rapidly sputters, introducing metal atoms and ions into the plasma for ISS conditions. The threshold for Ti atom ionization is much lower than ionization of argon atoms (6.8 eV for Ti atom vs. 15.75 eV for Ar atoms) causing a rapid increase in the electron/ion densities in the plasma for the ISS condition.

The emission spectrum can used to make a qualitative estimate of how the ratio of Ti ions to Ti atom concentration varies with input microwave powers. This approach has been used in similar studies for fractional ionization estimation [30-31]. For a same species plasma (e.g., only Ar plasma), the intensity ratio of two lines ($I_1/I_2 \approx (k_1/k_2)(N_1/N_2)$, where $k$ is the rate constant for excitation and $N$ is the concentration of the species) is proportional to the rate constant of excitation of the species, which reflects an electron mean energy in the plasma, i.e., the electron temperature [32]. For different species existing in an excited environment, the intensity ratio reflects their relative concentrations [32-33]. So, the intensity of a given emission line is proportional to the concentration of the species and hence the ratio of relative intensity of a Ti ion line to a Ti neutral line is proportional to the ionization ratio.

Figure 7 shows the comparison of emission spectra in the wavelength range of 425-480 nm for OS and ISS conditions (zoom of Fig. 6). As discussed above the Ti lines appear only for the ISS condition. Also, the Ar, Ti and O lines are observed for ISS only in this wavelength range. Here we choose the Ti ion line at 430.75 nm and Ti neutral line at 467.5 nm for light-emission intensity ratio. These lines are chosen on the basis of high intensity and low upper-level excitation energy. So we can measure the ionization fraction of sputtered Ti as a function of input microwave power.

Figure 8 shows the variation of emission intensity ratio of the Ti ion line to the Ti neutral line with axial positions for various microwave input powers $P_{in}$. The line intensity ratio is highest at the 0 mm axial location and drops rapid along the axis up to 50 mm. From 50 mm to about 220 mm the ratio remain nearly constant and then increases again as the microwave inlet is approached. Essentially, the low TiII/TiI ratio corresponds to the locations where the total ion densities are highest (see Fig. 5). An understanding of this trend can be obtained by observations of Hopwood et al. [34-35] in ionized vapor deposition where metal ions (Ti and Al) were observed in the entire plasma volume of their reactor, while the metal atoms are seen only at the target's surface where they originate as a result of sputtering processes. This is a consequence of sputtered atoms being rapidly ionized as soon as they enter the plasma. In our case, the entire antenna is at the same DC bias potential and the total ion densities are highest in the central region (50 to 220 mm). This implies the efflux of sputtered Ti atoms from the antenna is highest in the central region. The Ti atoms are rapidly ionized once they enter the plasma and subsequently transported to other regions of the plasma volume, i.e. towards the ends of the antenna. The result is that the ratio of the Ti ion to Ti atom (TiII/TiI) is highest at the ends of the antenna compared to the central region as evident from Fig. 8. The TiII/TiI line ratio increases over the entire length of the antenna with increasing $P_{in}$. This is clearly a consequence of the higher ionization fraction of the plasma with increasing power.

Another important point of discussion is the time interval over which the OS condition is sustained. As discussed earlier, the time duration of the OS must correspond to the thickness of the protective oxide layer on the antenna. We perform further studies to provide insights on this issue. Figure 9 shows the discharge current at the antenna as a function of discharge run time for a first 30 minute run that is followed by an off time of 120 minutes, following which the discharge was switched on again. For the first 30 minute discharge run, the plasma starts with OS conditions (low current) and transitions to steady ISS conditions (high current) after about 15 minutes. When the discharge is switched on again after the 120 minute off time, the plasma starts with incipient ISS conditions with lower currents and rapidly becomes a steady high current ISS. Presumably the off time of 120 minutes causes a gradual re-oxidization of the antenna surface and a lower antenna temperature due to cooling. For higher off time we therefore expect discharge will start with a more complete OS conditions followed by the transition to ISS. Further confirmation of the above picture is obtained for studies where the vacuum chamber is pressurized to atmospheric conditions during the off time. For such runs then the discharge starts plasma starts from the OS. Figure 3 also shows the OS sustaining time duration is 15 minutes for 30 W, which is more than for 40 W (13 min) and 50 W (10 min). For later times (ISS), the temperature of metal antenna is expected to be very high [36-37] due to ion bombardment and sputter yield is therefore expected to be even higher.

Overall the picture that emerges is that the presence of an oxide layer acts as a protective layer that prevents sputtering of the metal antenna. Even though the oxide layer is protective, it does experience a slow sputter erosion. The sputter yield of the oxide layer is also temperature dependent with increasing sputter yield with increasing temperature [36]. The oxide layer therefore erodes away more rapidly if the discharge starts with the antenna at a high temperature (shorter OS period).

A thick initial oxide layer should result in a longer OS period.

In the present MVP sources, in ISS, high density plasma is generated for relatively low input microwave powers, around 30 W. At ISS, the high density plasma is mainly due to Ar and sputtered Ti ions and a large diffusion of metal (Ti) ions throughout the vacuum chamber, which increases the plasma column size. Normally, in conventional sputtering cases, plasma source is non-uniform (high density plasma near to the source). But in MVP sources, nearly uniform high density plasmas are produced by very low input microwave powers (around 30 W, which is much less than the conventional sputtering microwave power, greater than 200 W). It is expected that if the antenna length will cover the whole chamber length, then more uniform plasma will be generated.

## IV. SUMMARY

In summary, two states of MVP are observed with time, as oxidized state and ionized sputter state. The OS sustaining time is dependent on oxidized layer thickness and input mw power. Length of plasma columns and ion density at ionized sputtered state are directly proportional to the input microwave powers. High density plasma (around $1\times10^{12}$ $cm^{-3}$) is generated for low input microwave power in such type of MVP sources, where the extended antenna plays an important role for microwave power coupling. The present experimental results suggest that MVP source is a best device for thin film deposition and ultra-high-speed DLC coating.

## ACKNOWLEDGEMENTS

This work was supported partly by DAIKO Foundation RESEARCH FELLOWSHIP PROGRAM in FY2014 and a “Grant for Advanced Industrial Technology Development (No. 11B06004d)” in 2011 from the New Energy and Industrial Technology Development Organization (NEDO) of Japan.

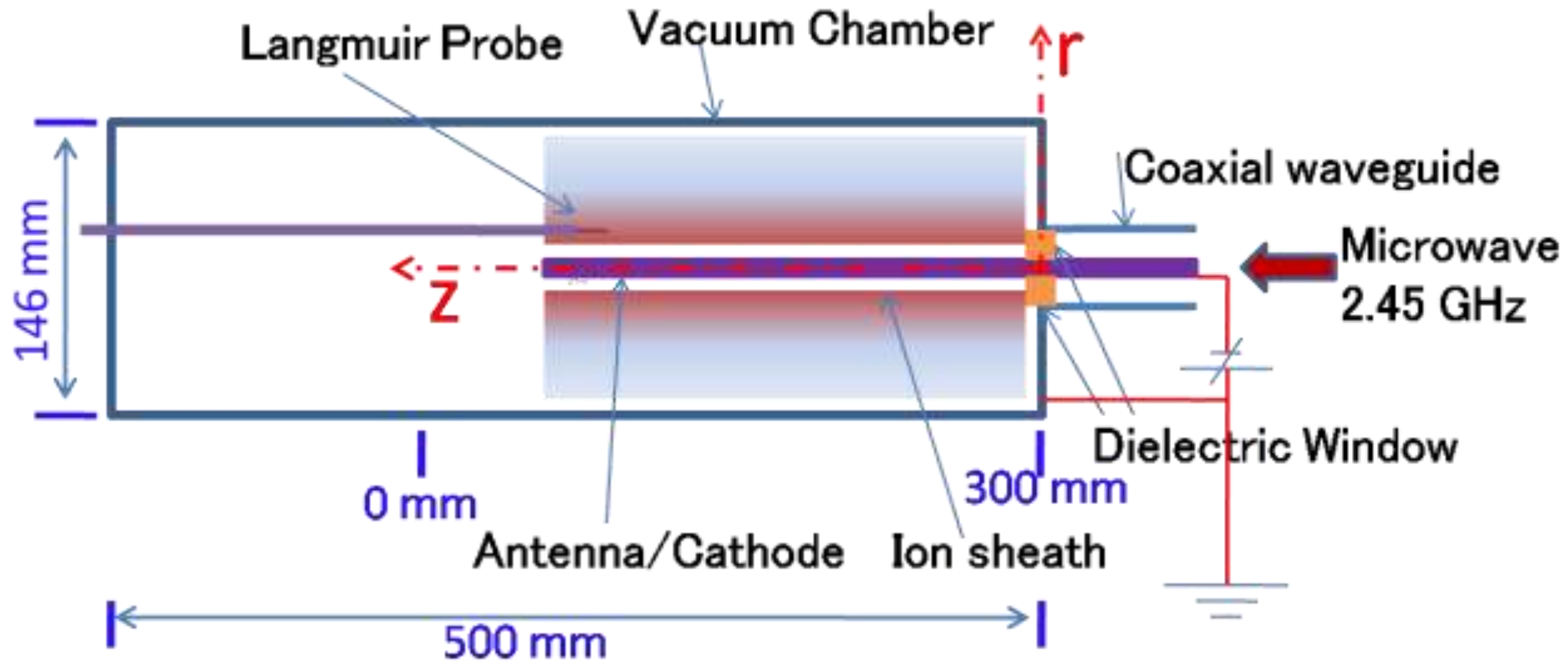

**Fig. 1.** Schematic of the experimental setup. Length of the vacuum chamber is 500 mm and inner diameter is 146 mm. Langmuir probe measurements are taken from left side (0 mm) of the vacuum chamber. 300 mm is the antenna zero position.

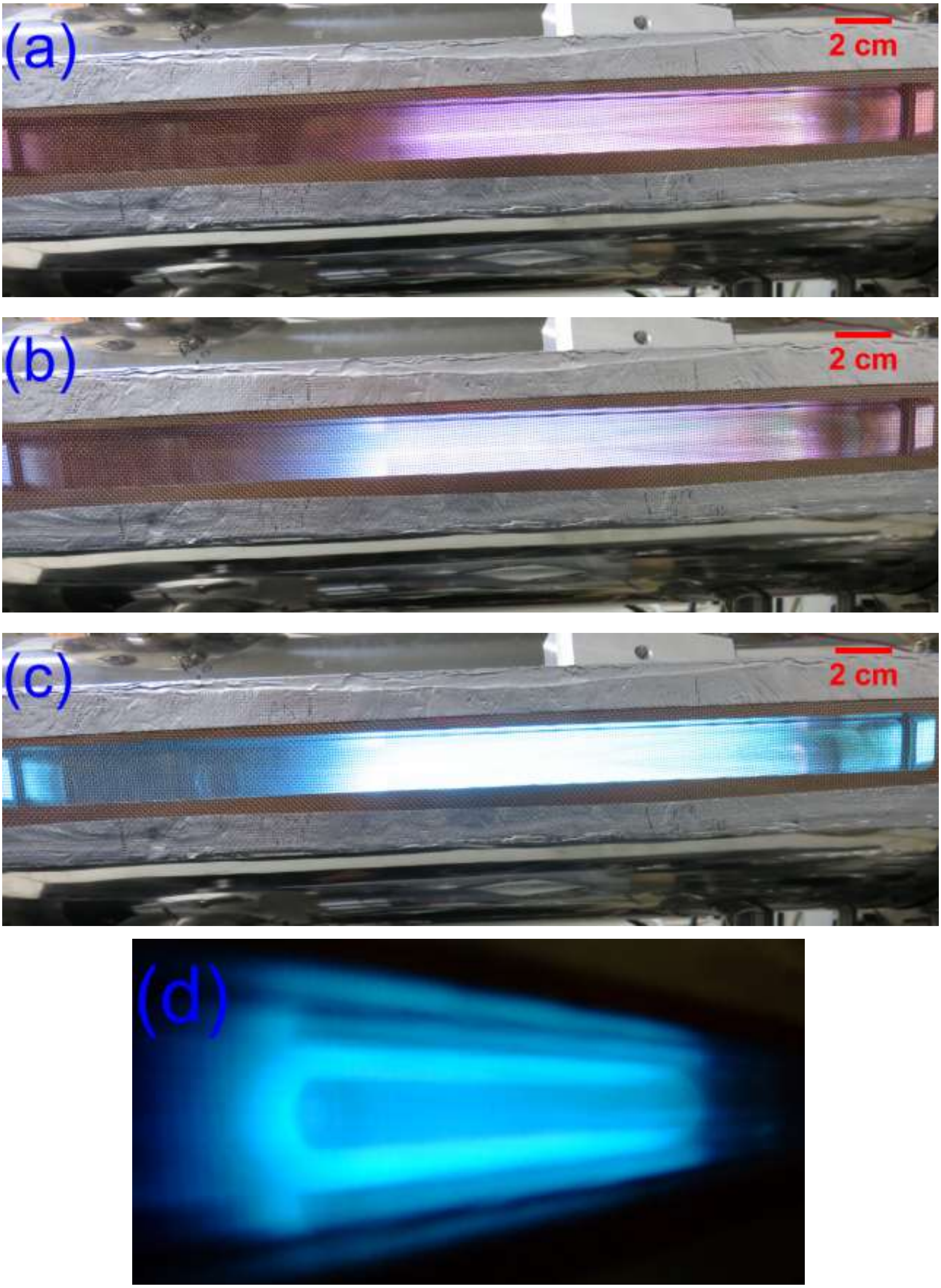

**Fig. 2.** Photographs of the plasma columns. (a) at 7 minutes, (b) at 11 minutes, (c) at 30 minutes and (d) close-up view at 30 minutes. The applied DC bias (-300 V), the gas pressure (50 Pa) and $P_{in}$ (50 W) were fixed.

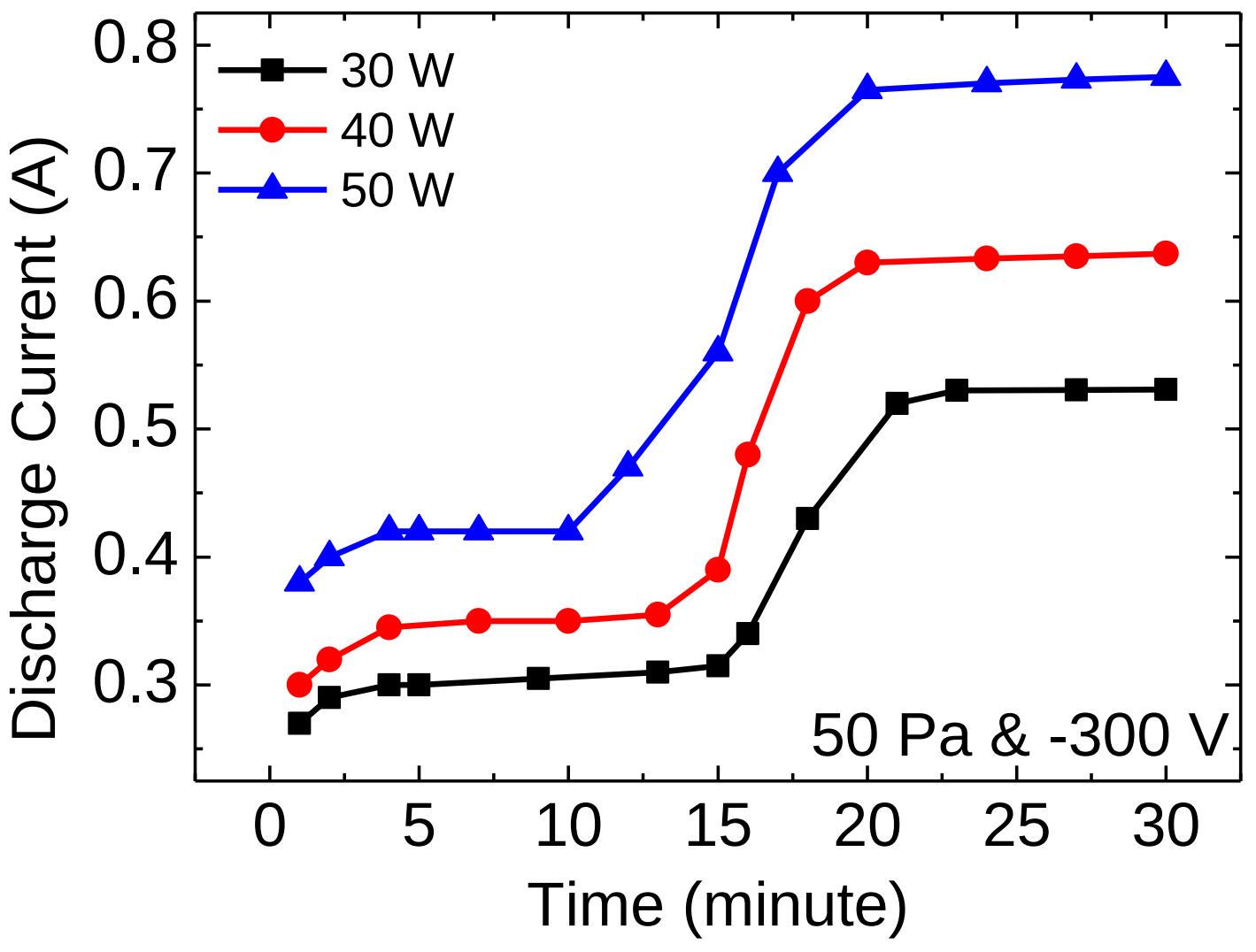

**Fig. 3.** Variation of the plasma discharge current at the DC biased antenna with time for various input microwave powers.

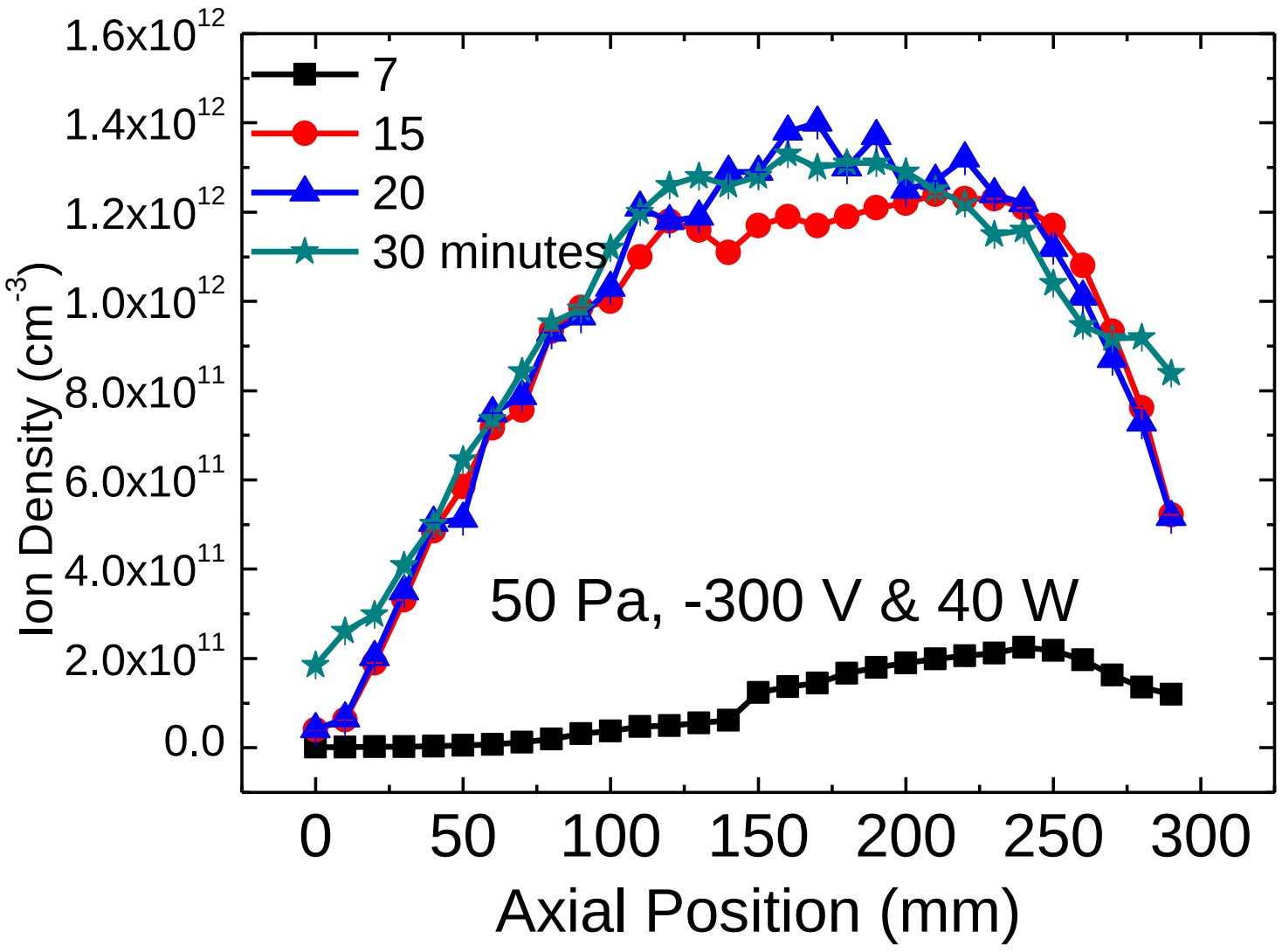

**Fig. 4** Axial distribution of ion density for OS (7 minutes) and ISS (15, 20 and 30 minutes). Here the applied DC bias (-300 V), gas pressure (50 Pa) and $P_{in}$ (40 W) were fixed. (Note the axial positions 0 to 300 mm is described in Fig. 1)

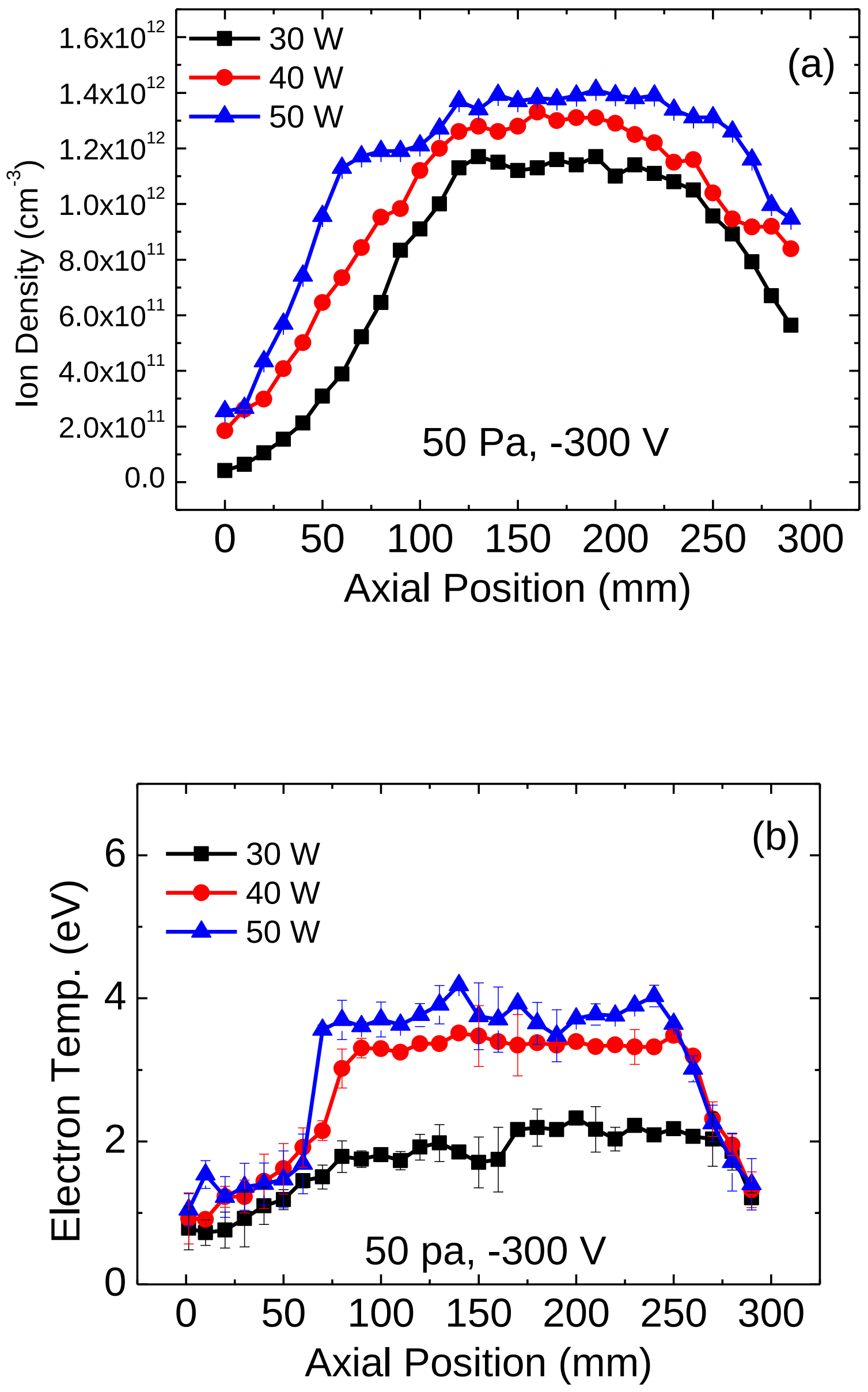

**Fig. 5.** Axial distribution of (a) ion density and (b) electron temperature for various microwave input powers $P_{in}$ under ISS conditions. Here the applied DC bias (-300 V) and gas pressure (50 Pa) were fixed.

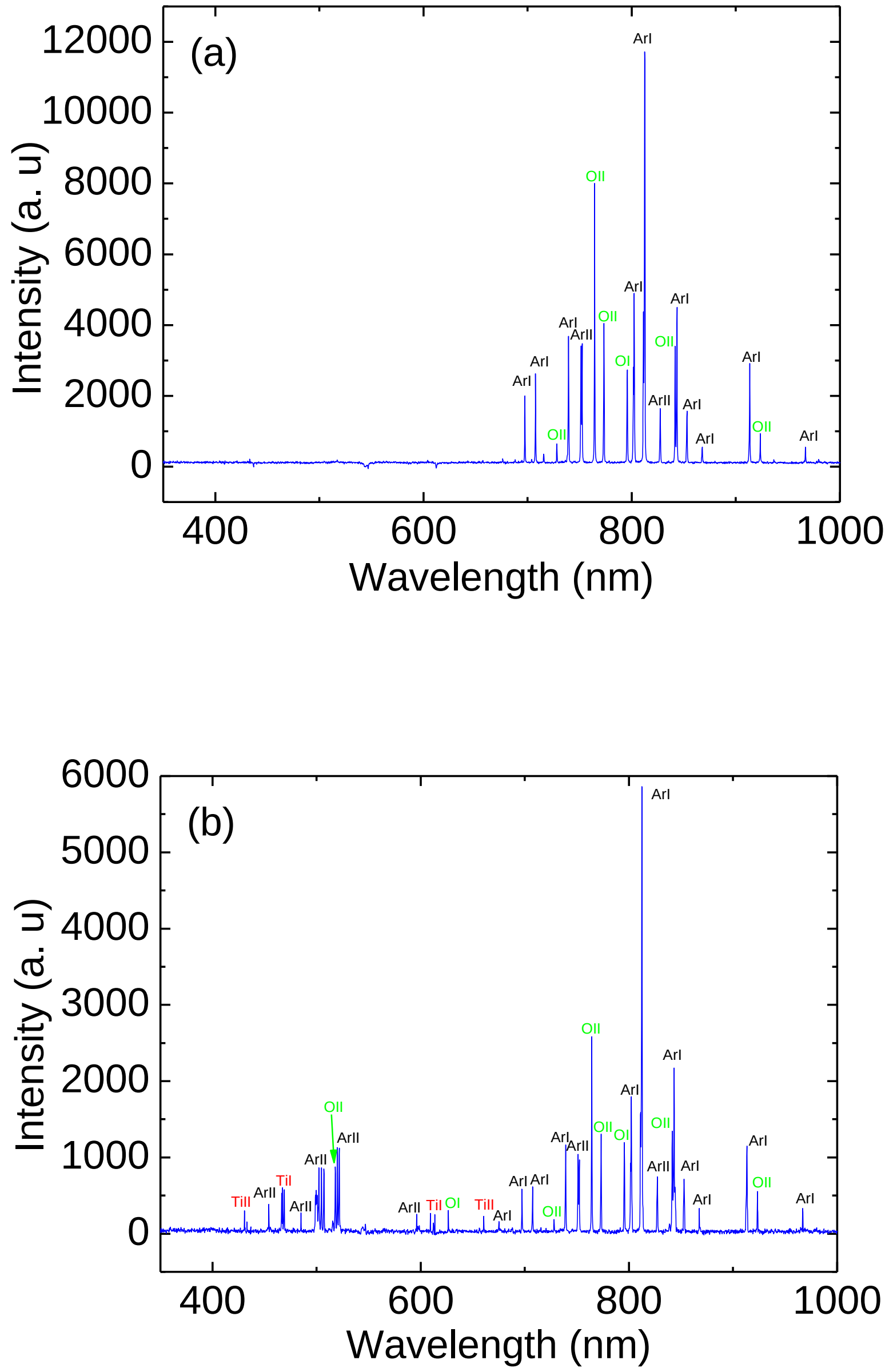

**Fig. 6.** Typical emission spectrum (a) for OS (7 minutes) and (b) for ISS (30 minutes). Here the input microwave power (30 W), applied DC bias (-300 V) and the gas pressure (50 Pa) were fixed. The axial position is 120 mm away from the antenna zero position.

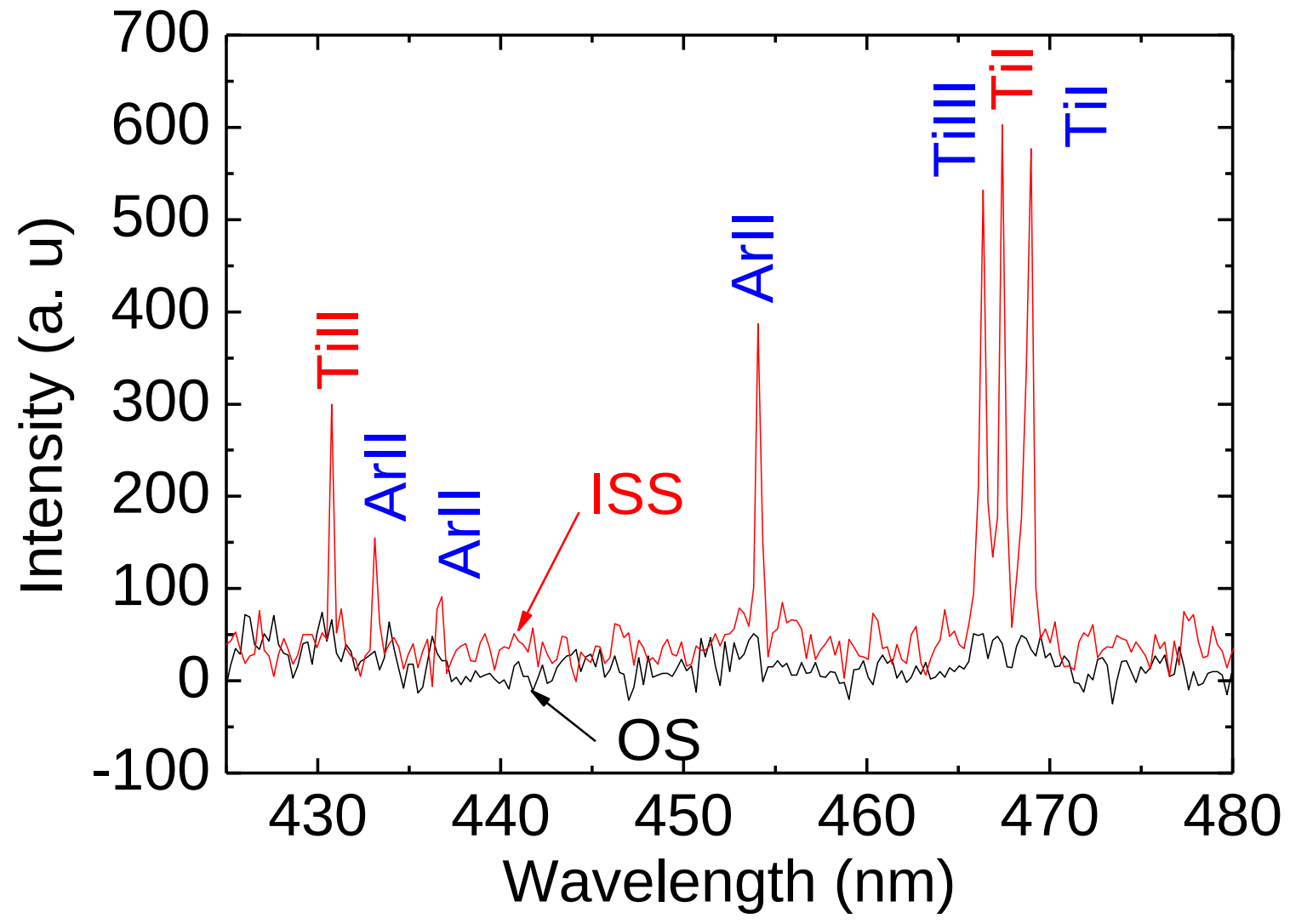

**Fig. 7.** Emission spectra for OS and ISS conditions in the wavelength range of 425-480 nm (Zoom of Fig. 6). Ti lines are seen for ISS conditions.

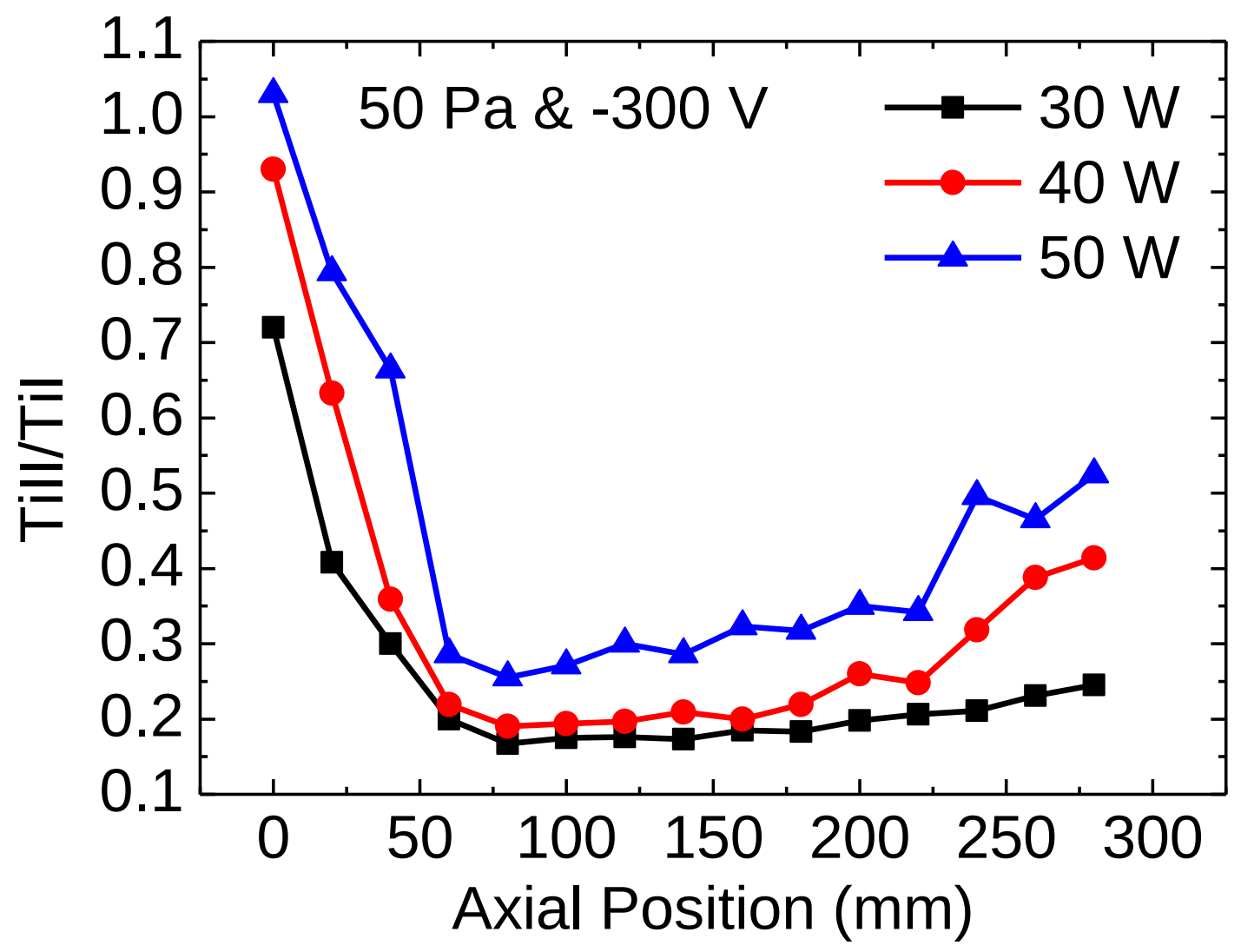

**Fig. 8.** Emission intensity ratio of Ti ion line (430.75 nm) to Ti neutral line (467.5 nm) for ISS condition as a function of axial positions at various input microwave powers.

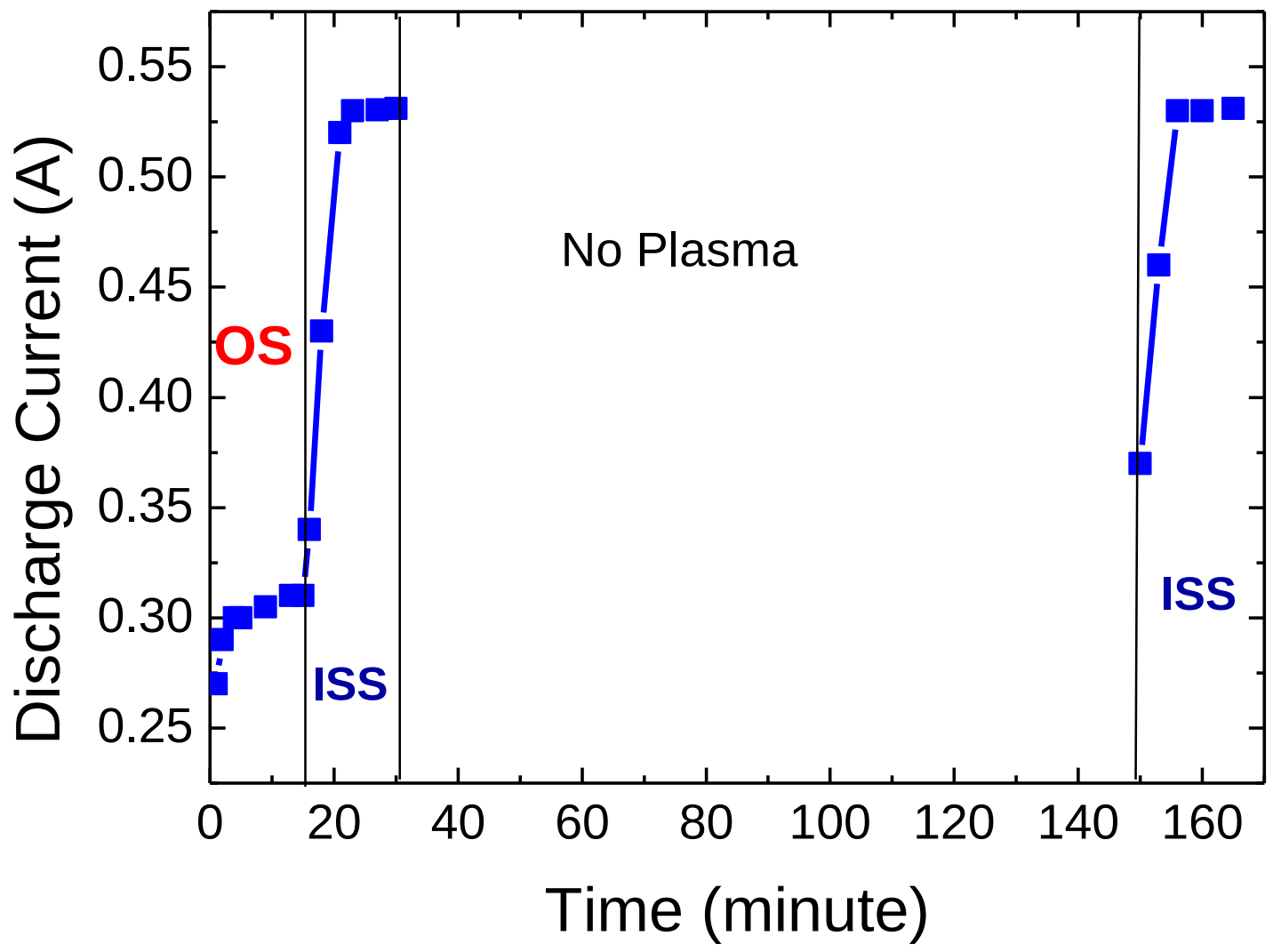

**Fig. 9.** Variation of discharge currents at antenna with time. Here the input microwave power (30 W), applied DC bias (-300 V) and the gas pressure (50 Pa) were fixed.